\documentstyle{l-aa}
\begin{document}

\title{Comment on a paper by M.A. Bautista ``Atomic data from the Iron Project. 
XVI. Photoionization cross sections and oscillator strengths for Fe V''}
\author{
D.A.~Verner\inst{1}
\and
D.G.~Yakovlev\inst{2}
\and
I.M.~Band\inst{3}
\and
M.B.~Trzhaskovskaya\inst{3}
}
\institute{
Department of Physics and Astronomy, University of Kentucky,
Lexington, KY 40506, U.S.A.\\Internet: verner@pa.uky.edu
\and
Ioffe Physical Technical Institute,
St.~Petersburg 194021, Russia
\and
Nuclear Physics Institute, Gatchina, St.~Petersburg 188350, Russia
}
\thesaurus{02.01.3, 02.01.4}
\date{Received date; accepted date}
\maketitle

\label{sampout}

\begin{abstract}
Recently, Bautista (1996) reported new
calculations of photoionization cross sections for Fe~{\sc v}, and
compared them with the earlier results of Reilman \& Manson
(1979) and Verner et al. (1993). Bautista claimed that beyond 10~Ry the new
cross sections ``converge well toward the results by Reilman and
Manson'', whereas ``the calculations by Verner et al. still
underestimate the cross section by almost factor of two''.
We show that Bautista erroneously compared the total (summed over
shells) cross sections from Reilman \& Manson with the partial
$3d$-shell cross sections from Verner et al.
Actually, the total cross sections
from Verner et al. and Reilman \& Manson agree within 3\%
at all energies.
\end{abstract}

In a recent paper, Bautista (1996) presented new R-matrix
calculations of photoionization cross sections for Fe~{\sc v}, and
compared them with the Hartree-Slater data of Reilman \& Manson
(1979, hereafter RM) and Hartree-Dirac-Slater data of Verner et al.
(1993, hereafter VYBT). In his paper, Bautista wrote:

``Beyond 10 Ry the present cross sections converge well toward the results
by Reilman and Manson as might be expected since the electron correlation
effects get weaker with increasing energy. However, the calculations by
Verner et al. still underestimate the cross section by almost factor of two.
The low values of the photoionization cross section from Verner et al. with
respect to those by Reilman and Manson are not understood since both of
these calculations are based on a similar approximation.''

The reason of this ``disagreement'' is very simple. RM
published the tables of total (summed over shells) cross sections.
VYBT presented analytic fits to the partial 
cross sections for separate shells.
Bautista erroneously compared the total
cross section from RM with the partial
$3d$-shell cross section from VYBT.
The total cross sections from VYBT and RM agree within
relative error $\delta < 3\%$
at all photon energies $E$.
To illustrate that, we present (Table 1)
the photoionization
cross sections of Fe~{\sc v} calculated with the VYBT analytic fits for the
first 10 entries of the RM table. Note that the $3d$-shell
ionization threshold of Fe~{\sc v} is 73.03 eV (5.37 Ry), the $3p$-shell
threshold 128.8 eV (9.47 Ry), and the $3s$-shell threshold 163.3 eV (12.0 Ry)
(see VYBT).

\begin{table}
\caption{Photoionization cross section $\sigma$ [Mb] of Fe V}
\begin{tabular}{r|cccc|c|c}
\hline
 & \multicolumn{4}{c|}{VYBT} & RM & \\
E, eV & $3d$ & $3p$ & $3s$ & Total & Total & $\delta$, \%\\
\hline
80  & 5.341 & 0     & 0     & 5.341 & 5.350 & 0.2\\
90  & 4.788 & 0     & 0     & 4.788 & 4.779 & 0.2\\
100 & 4.190 & 0     & 0     & 4.190 & 4.204 & 0.3\\
130 & 2.694 & 1.073 & 0     & 3.767 & 3.855 & 2.3\\
160 & 1.750 & 1.121 & 0     & 2.871 & 2.919 & 1.6\\
190 & 1.176 & 1.043 & 0.271 & 2.490 & 2.531 & 1.6\\    
210 & 0.920 & 0.968 & 0.244 & 2.132 & 2.165 & 1.5\\
240 & 0.653 & 0.851 & 0.210 & 1.714 & 1.736 & 1.3\\
270 & 0.478 & 0.742 & 0.182 & 1.402 & 1.416 & 1.0\\
300 & 0.358 & 0.647 & 0.158 & 1.163 & 1.172 & 0.8\\
\hline
\end{tabular}
\end{table}

\end{document}